\documentclass{PoS}
\usepackage{graphicx}
\usepackage{amsmath}
\usepackage{slashed}
\newcommand{\helacdipoles}{\texttt{HELAC-Dipoles}}
\newcommand{\helaconeloop}{\texttt{HELAC-Oneloop}}
\newcommand{\helacphegas}{\texttt{HELAC-PHEGAS}}

\newcommand{\pythia}{\texttt{PYTHIA}}
\newcommand{\herwig}{\texttt{HERWIG}}

\newcommand{\madloop}{\texttt{MADLOOP}}
\newcommand{\powheghelac}{\texttt{POWHEG+HELAC}}
\newcommand{\powhel}{\texttt{PowHel}}
\newcommand{\powhegbox}{\texttt{POWHEG-BOX}}
\newcommand{\fastjet}{\texttt{FastJet}}

\newcommand{\ttjet}{t\,$\bar{{\rm t}}$ + jet }
\newcommand{\ttH}{t\,$\bar{{\rm t}} + H$ }
\newcommand{\ttX}{t\,$\bar{{\rm t}} + X$ }

\newcommand{\HT}{\ensuremath{H_{\perp}}}

\newcommand{\bt}{\ensuremath{\bar{{\rm t}}}}

\newcommand{\NLO}{{\rm NLO}}
\newcommand\fig[1]     {Fig.\,{\ref{#1}}}

\title{NLO event samples for the LHC}

\ShortTitle{NLO event samples for the LHC}

\author{M.V.~GARZELLI, Adam KARDOS and
\speaker{Zolt\'an TR\'OCS\'ANYI}%
\thanks{This research was supported by
the LHCPhenoNet network PITN-GA-2010-264564,
and the T\'AMOP 4.2.1./B-09/1/KONV-2010-0007 project.}\\
Institute of Physics, University of Debrecen,
H-4010 Debrecen P.O.Box 105, Hungary \\
       E-mail: \email{Z.Trocsanyi@atomki.hu}}

\abstract{
We introduce a twiki page with collections of generated Monte Carlo
event samples in proton-proton collisions at LHC energies including a
heavy quark-antiquark pair in the final state. These samples
are generated with the POWHEG method and can be used to prepare
distributions at the NLO accuracy with first radiation treated
according to the parton shower approach. Information related to each
event is stored in the form prescribed by the Les Houches Accords.
Standard parton shower Monte Carlo programs can be used to further
evolve these events, and simulate events at the hadron level, ready for
almost arbitrary experimental analysis. Currently the available final
states are the following: (i) t + \bt, (ii) t + \bt\ + H,
(iii) 
t + \bt\ + jet, while the generation of several other final states
is in progress.
}

\FullConference{XXIst International Europhysics Conference on High Energy
Physics\\
		 21-27 July 2011\\
		 Grenoble, Rhones Alpes France}

\begin{document}



In recent years a lot of NLO QCD calculations have been presented in
the literature. High-energy experiments, notably at the LHC, have and
will be benefited by the progress in our computational ability to deal
with higher order corrections in scattering amplitudes with many
partons involved. In order though to get the optimum benefit and to
produce predictions that can be directly compared to experimental data
at the hadron level, a matching with parton showers (PS) and
hadronization is ultimately inevitable. 

In order to construct a generic interface between parton showers and
matrix element calculations at the NLO accuracy, we have chosen to
combine the HELAC-NLO~\cite{Bevilacqua:2010mx,Bevilacqua:2011xh} and POWHEG
\cite{Nason:2004rx,Frixione:2007vw} approach respectively. Our goal is
to apply our method to the hadroproduction of a t\bt\ pair in association
with Standard Model boson(s) or fermion(s) (denoted X). Due to their
largest mass, the t-quarks play an outstanding role in the Standard
Model and top physics is expected to yield a lot of new experimental
information at the LHC (see contributions in this proceedings).  The
first applications of our project were the processes $pp \to$ \ttjet\
\cite{Kardos:2011qa} and $pp \to$ \ttH\ \cite{Garzelli:2011vp}.  In
this contribution we introduce a web-page that is planned to contain
event samples in the form of the Les Houches accord
\cite{Alwall:2006yp} for such processes.  These events are readily
usable for producing distributions at the hadron level including NLO
QCD accuracy for the hard scattering process, and we expect that they
will be very useful for experimental analyses.  


We performed our calculations using the \powhegbox\ \cite{Alioli:2010xd},
which requires the following ingredients:
1) flavor structures of the Born and real radiation subprocesses,
2) Born-level phase space,
3) squared matrix elements with all incoming momenta for the Born and the
real-emission processes are built using amplitudes obtained from
\helaconeloop\ \cite{vanHameren:2009dr} and \helacphegas\
\cite{Cafarella:2007pc}, respectively.  The matrix elements in the
physical channels are obtained by crossing,
4) color-correlated squared matrix elements are taken from
\helacdipoles\ \cite{Czakon:2009ss},
5) we use the polarization vectors to project the helicity-correlated
amplitudes to Lorentz basis for writing the spin-correlated squared
matrix elements.
With this input \powhegbox\ can be used to generate events at the Born
level plus first radiation and store those in Les Houches Event Files
(LHEF). Then one can choose any parton shower (PS) Monte Carlo program
for generating events with hadrons.

In the \powhegbox\ the first emission is the hardest one measured by
transverse momentum which is can also be chosen the ordering variable
in \pythia.  If the ordering variable in the shower is different from
the transverse momentum of the parton splitting, such as in \herwig,
then the hardest emission is not necessarily the first one. In such
cases \herwig\ discards shower evolutions (vetoed shower) with larger
transverse momentum in all splitting occuring after the first emission.
In addition, a truncated shower simulating wide-angle soft emission
before the first emission is also needed in principle, but its effect
was found small \cite{LatundeDada:2006gx,Kardos:2011qa,Garzelli:2011vp}.
As there is no implementation of truncated shower in \herwig\ using
external LHE event files, the effect of the truncated showers is absent
from our predictions.


As default, we check the consistency between real-emission, Born,
color-correlated and spin-correlated matrix elements in randomly chosen
phase space regions by taking the soft and collinear limits of the
real-emission matrix elements in all possible kinematically
degenerate channels. We compare the virtual contributions to the
predictions of other programs for computing one-loop amplitudes, such
as \madloop\ \cite{Hirschi:2011pa}, if available. In all cases we
found agreement up to 5--6 digits.  

\begin{figure}
\includegraphics[width=0.49\linewidth]{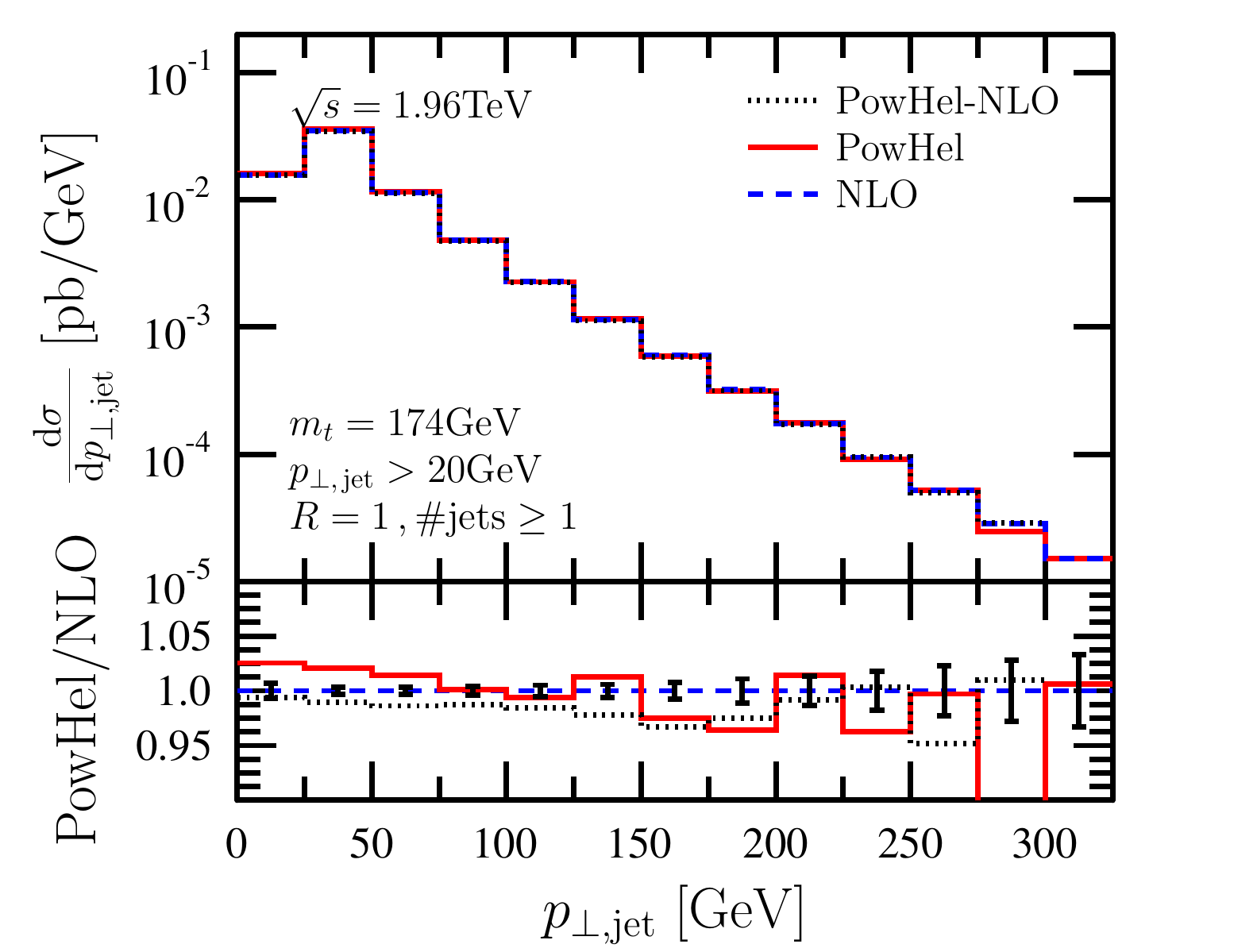}
\includegraphics[width=0.49\linewidth]{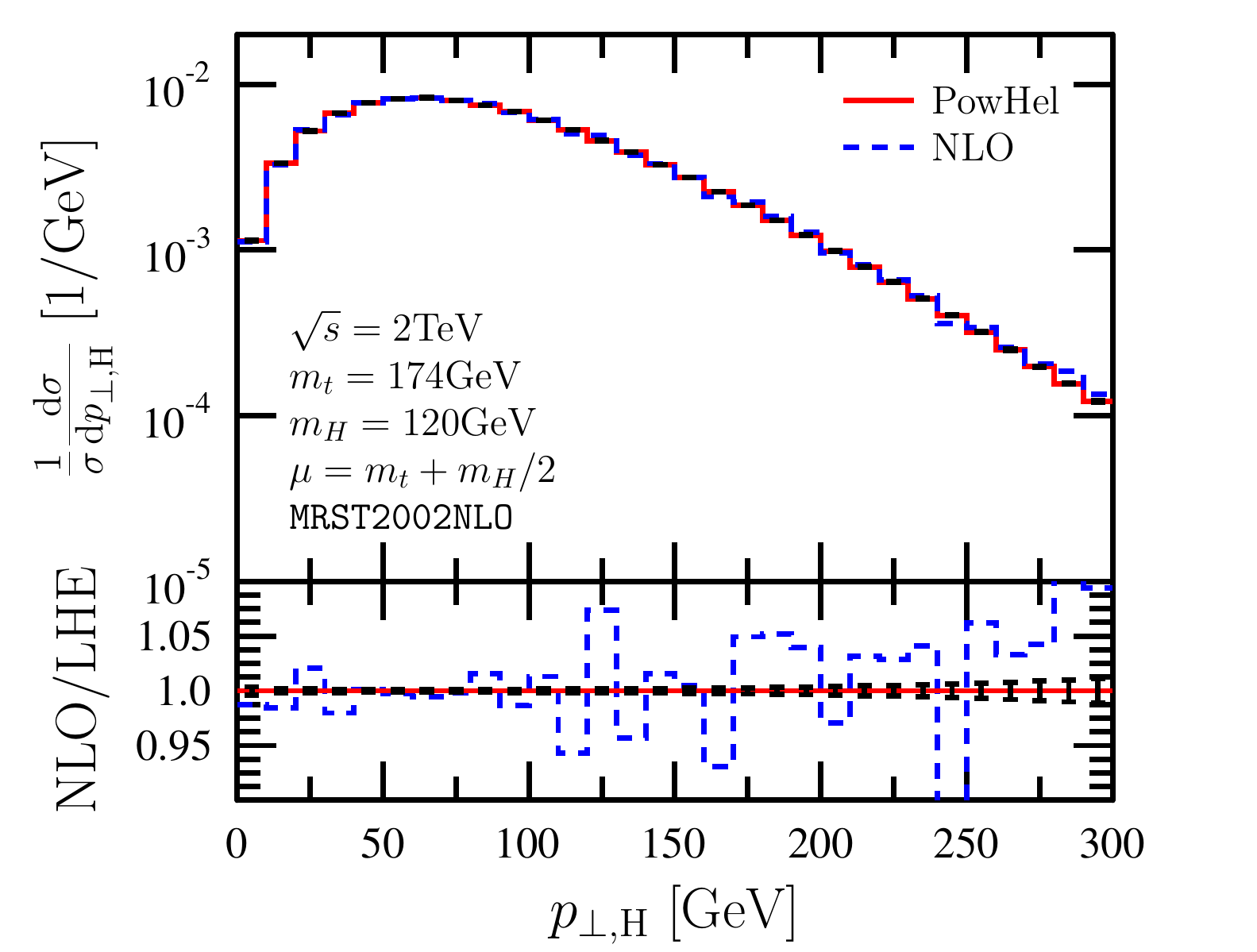}
\includegraphics[width=0.49\linewidth]{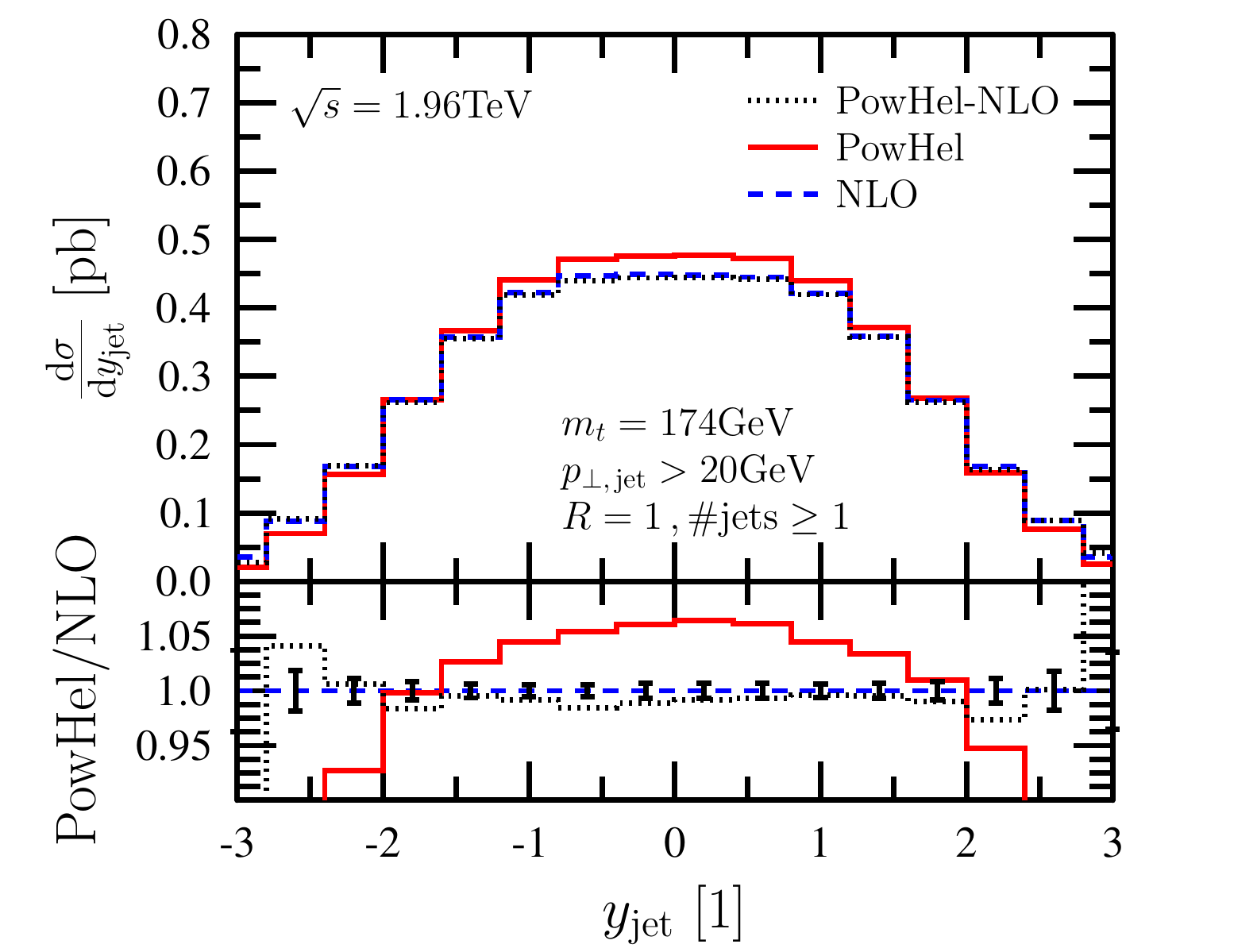}
\includegraphics[width=0.49\linewidth]{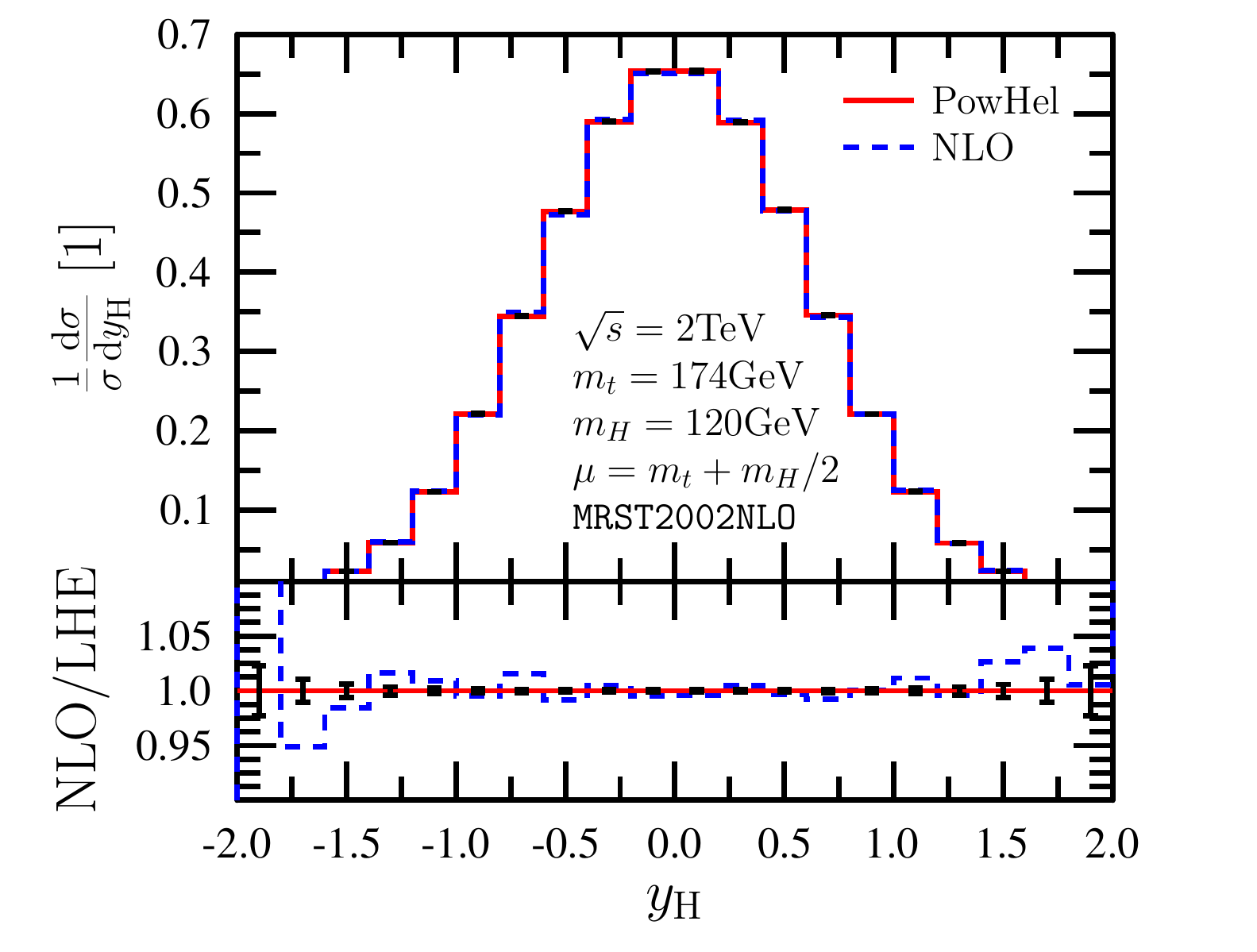}
\caption{Distributions of transverse momemtum (upper panels)
and rapidity (lower panels) of the jet in the process 
$pp \to$ \ttjet\ (left) and of the Higgs particle in the process
$pp \to$ \ttH\ (right).
The insets below each plot show the ratio of the predictions (PowHel/NLO).}
\label{fig:POWHEGvsNLOdistributions}
\end{figure}

The processes we considered were also studied extensively at the NLO
accuracy in the literature, which enables us to make detailed checks of
our predictions.  For each process we compared the cross sections at
the \NLO\ accuracy to already published predictions, if available, and
found agreement within the statistical uncertainty of our integrations.
Finally, we also compared differential distributions based on events
already including first radiation from the \powhegbox\ with the
corresponding \NLO\ predictions.  As examples, we show transverse
momentum and rapidity distributions in
\fig{fig:POWHEGvsNLOdistributions} for the relevant processes. The
difference between the NLO and \powhel\ (\powheghelac) predictions are
expected to be beyond NLO accuracy \cite{Nason:2004rx}. 

Next we present selected predictions \ttX\ production with parton
shower and hadronization effects at LHC. We used the event files from
our webpage. Together with each set of events, one can download a
sample analyis file to produce the plots presented here. In each plot we
include our selection cuts, which can be altered almost freely to perform
different analyses on the same events. Also the set of physical parameters
used during the generation of the events can be found together with the
event files.  We used the last version of the PS programs, \pythia~6.425
\cite{Sjostrand:2006za} and \herwig~6.520 \cite{Corcella:2002jc}, as
well as for jet reconstruction, \fastjet~2.4.3 \cite{Cacciari:2005hq}.  

\begin{figure}
\includegraphics[width=0.49\linewidth]{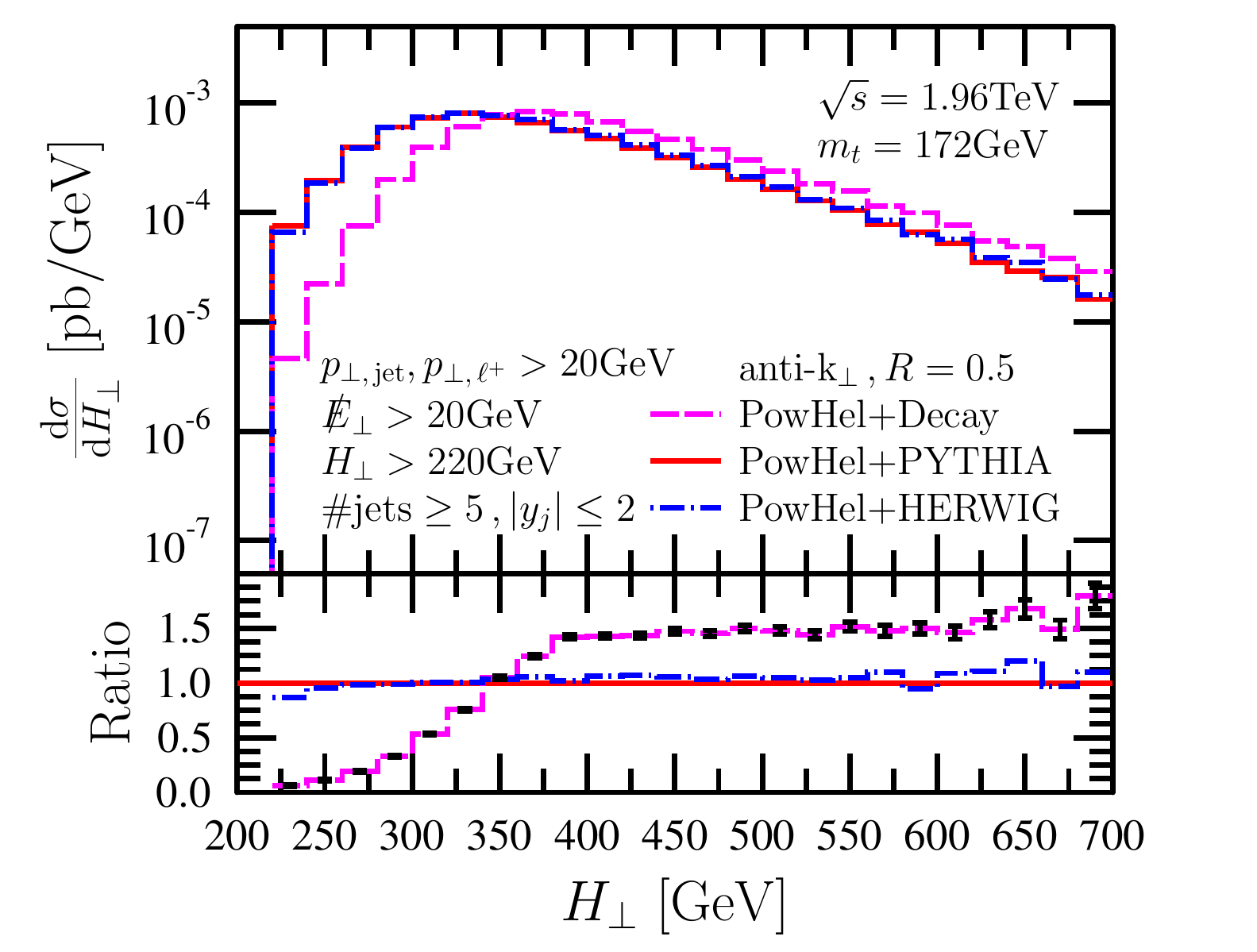}
\includegraphics[width=0.49\linewidth]{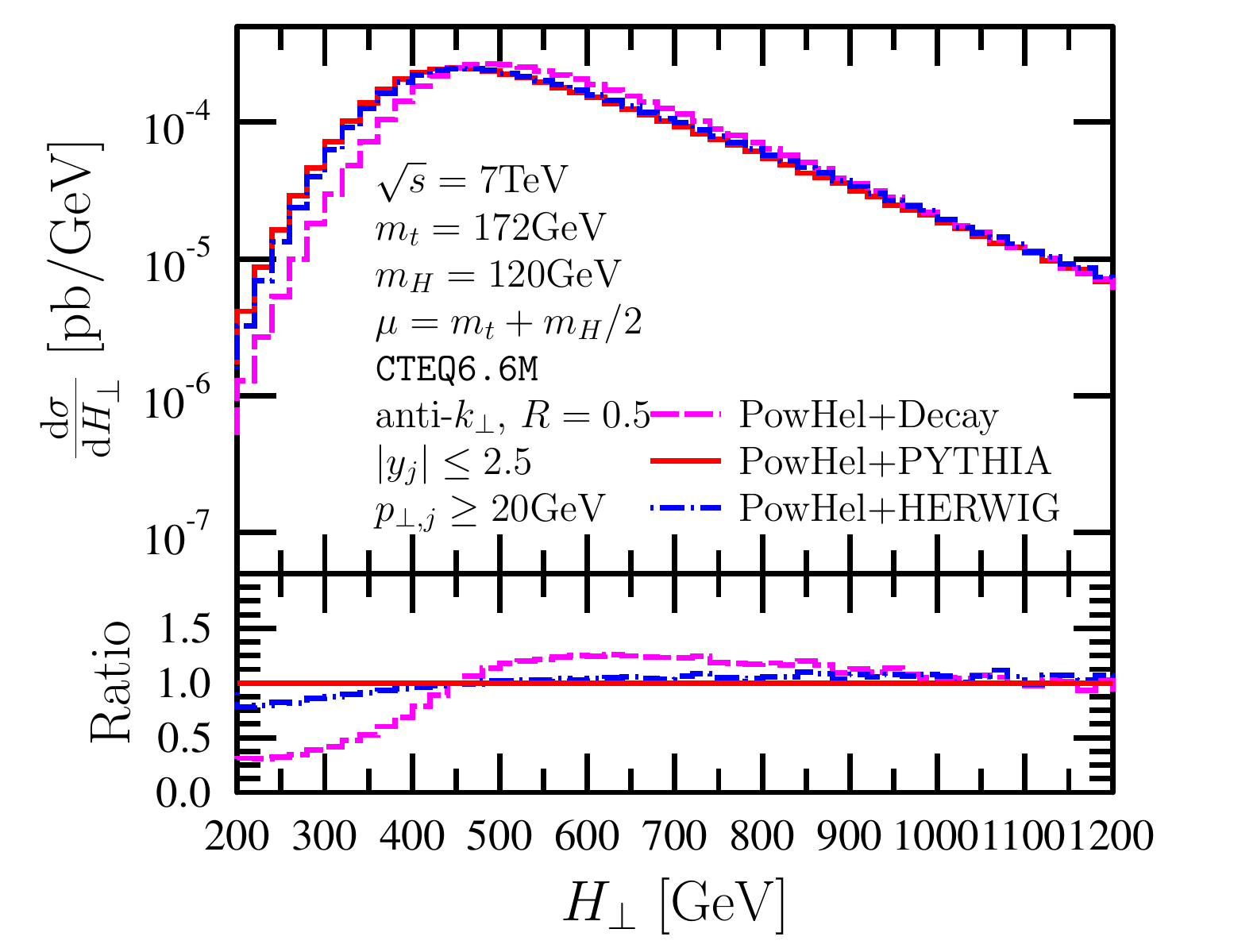}
\caption{\HT-distributions of the
$pp \to$ \ttjet\ (left) and of $pp \to$ \ttH\ (right) processes after
decay of the heavy particles and after full shower Monte Carlo.
The lower panels show the ratio of the predictions at the decay level 
to the shower MC level (\powhel+\pythia) and the ratio of the
predictions at the hadron level using different shower MC codes
(\powhel+\herwig/\powhel+\pythia).}
\label{fig:ht}
\end{figure}

In \fig{fig:ht} we present distributions of the scalar sum of
transverse momenta in the event, \HT for the two processes $pp \to$
\ttjet\ and $pp \to$ \ttH. The selection cuts implemented for these
analyses are specified in Refs.\cite{Kardos:2011qa} and
\cite{Garzelli:2011vp}, respectively.  We can observe a softening of
the spectra, when the shower is turned on, as expected on the basis of
the parton shower approach.


Interfacing NLO calculations, as structured in HELAC-NLO, with PS and
hadronization effects, within the POWHEG framework, open the exciting
possibility of realistic, precise and reliable simulations. We forsee
large experimental and phenomenological potential of our approach. The
event files produced by the \powhegbox\ for processes $pp \to$ t\bt+X,
where $X$ is a hard object (SM boson, jet etc), together with the
corresponding version of the program are available at
http://grid.kfki.hu/twiki/bin/view/DbTheory/TthProd.



\begin{thebibliography}{99}
\itemsep=-2pt
\bibitem{Nason:2004rx}
  P.~Nason,
  JHEP {\bf 0411}, 040 (2004)
  [arXiv:hep-ph/0409146].

\bibitem{Frixione:2007vw}
  S.~Frixione, P.~Nason and C.~Oleari,
  JHEP {\bf 0711}, 070 (2007)
  [arXiv:0709.2092].

\bibitem{Bevilacqua:2010mx}
  G.~Bevilacqua {\it et al.}.
  Nucl.\ Phys.\ Proc.\ Suppl.\  {\bf 205-206}, 211 (2010)
  [arXiv:1007.4918].

\bibitem{Bevilacqua:2011xh}
  G.~Bevilacqua {\it et al.},
  arXiv:1110.1499.

%
\bibitem{Kardos:2011qa}
  A.~Kardos, C.G.~Papadopoulos and Z.~Trocsanyi,
  Phys.\ Lett.\ B {\bf 705}, 76 (2011)
  [arXiv:1101.2672].

\bibitem{Garzelli:2011vp}
  M.~V.~Garzelli, A.~Kardos, C.~G.~Papadopoulos and Z.~Trocsanyi,
  Europhys.\ Lett.\  {\bf 96}, 11001 (2011)
  [arXiv:1108.0387].

\bibitem{Alwall:2006yp}
  J.~Alwall {\it et al.},
  Comput.\ Phys.\ Commun.\  {\bf 176}, 300 (2007)
  [arXiv:hep-ph/0609017].

\bibitem{Alioli:2010xd}
  S.~Alioli, P.~Nason, C.~Oleari and E.~Re,
  JHEP {\bf 1006}, 043 (2010)
  [arXiv:1002.2581].

\bibitem{vanHameren:2009dr}
  A.~van Hameren, C.~G.~Papadopoulos and R.~Pittau,
  JHEP {\bf 0909}, 106 (2009)
  [arXiv:0903.4665].

\bibitem{Cafarella:2007pc}
  A.~Cafarella, C.~G.~Papadopoulos and M.~Worek,
  Comput.\ Phys.\ Commun.\  {\bf 180}, 1941 (2009)
  [arXiv:0710.2427].

\bibitem{Czakon:2009ss}
  M.~Czakon, C.~G.~Papadopoulos and M.~Worek,
  JHEP {\bf 0908}, 085 (2009)
  [arXiv:0905.0883].

\bibitem{LatundeDada:2006gx}
O.~Latunde-Dada, S.~Gieseke, B.~Webber,
  JHEP {\bf 0702}, 051 (2007)
  [arXiv:hep-ph/0612281].

\bibitem{Hirschi:2011pa}
  V.~Hirschi, R.~Frederix, S.~Frixione, M.~V.~Garzelli, F.~Maltoni and
R.~Pittau,
  JHEP {\bf 1105}, 044 (2011)
  [arXiv:1103.0621].

\bibitem{Sjostrand:2006za}
  T.~Sjostrand, S.~Mrenna and P.~Z.~Skands,
  JHEP {\bf 0605}, 026 (2006)
  [arXiv:hep-ph/0603175].

\bibitem{Corcella:2002jc}
  G.~Corcella {\it et al.},
  arXiv:hep-ph/0210213.

\bibitem{Cacciari:2005hq}
  M.~Cacciari and G.~P.~Salam,
  Phys.\ Lett.\  B {\bf 641}, 57 (2006)
  [arXiv:hep-ph/0512210], http://fastjet.fr

\end{thebibliography}
\end{document}